\documentclass[preprint,aip]{revtex4-1}
\usepackage{graphicx}
\usepackage{dcolumn}
\usepackage{bm}
\usepackage{amsfonts,amssymb,amsmath,geometry, url,siunitx}
\usepackage{gensymb}
\usepackage{graphicx}
\usepackage{hyperref}
\usepackage{setspace}
\usepackage[left]{lineno}
\usepackage{hyperref}
\hypersetup{
   colorlinks=true,    
   linkcolor=blue,       
   citecolor=blue,       
   filecolor=black,        
   urlcolor=blue,        
   linktoc=page            
}

\begin{document}

\title{Edge enhanced growth induced shape transition in the formation of GaN Nanowall Network
} %

\author{Sanjay  Nayak}
\email{sanjaynayak@jncasr.ac.in}
\affiliation{Chemistry and Physics of Materials Unit\\Jawaharlal Nehru Centre for Advanced Scientific Research (JNCASR), Bangalore-560064}
\author{Rajendra Kumar}
 \affiliation{Chemistry and Physics of Materials Unit\\Jawaharlal Nehru Centre for Advanced Scientific Research (JNCASR), Bangalore-560064}
\author{S.M. Shivaprasad}
\email{smsprasad@jncasr.ac.in}
\affiliation{Chemistry and Physics of Materials Unit\\Jawaharlal Nehru Centre for Advanced Scientific Research (JNCASR), Bangalore-560064}
\begin {abstract}
 We address the mechanism of early stages of growth and shape transition of the unique nanowall network (NwN) nanostructure of GaN by experimentally monitoring  its controlled growth using PA-MBE and complementing it by \textit{first-principles} calculations. Using electron microscopy, we observe the formation of tetrahedron shaped (3 faced pyramid) islands at early stages of growth, which later grows anisotropically along their edges of the (20$\overline{2}$1) facets, to form the wall like structure. The mechanism of this crystal growth is discussed in light of surface free energies of the different surfaces, adsorption energy and diffusion barrier of Ga ad-atoms on the (20$\overline{2}$1) facets. By \textit{first-principles} calculations, we find that the diffusion barrier of ad-atoms  decreases with decreasing width of facets, and is responsible for the anisotropic growth and formation of the nanowall network. This study suggest that formation of NwN is a archetype example of structure dependent attachment kinetic (SDAK) instability induced shape transition in thin film growth.
\end {abstract}

\pacs{ }

\maketitle

\section{Introduction}
The  dynamics  of  shape  transitions plays a critical role in the evolution of surface morphology and uniformity of hetero-epitaxial thin film and nanostructure growth.\cite{rastelli2001reversible,barth2005engineering,tersoff1993shape,brune1998self,consonni2010nucleation}. Critical to success of such nanostructure, especially while it is forming spontaneously, is the flexibility of the structure. However, lack of understanding on the role of attachment kinetics of adatoms and thermodynamic factors to the self assembly and the shape transitions is the major obstacle in achieving control over the growth of such structures. Therefore, investigations of the growth process of nanostructures enable us, to gain high control over the self-assembly. Further, the evolution of a structure from the early stages to the final morphology it attains is a very interesting and complex phenomena as many physical parameters compete with each other \cite{shchukin1999spontaneous,li2014thermodynamic,pimpinelli1998physics}. 
\par
III-Nitrides (InN, GaN and AlN) are  very special within semiconductors because of their applications in many areas due to direct and tunable band gap, mechanical and chemical stability, etc\cite{sang2013comprehensive}. Due to lack of availability of suitable native substrate, III-nitride films are typically grown on foreign substrates such as Sapphire, Silicon and Silicon carbide which results in formation of large density of defects\cite{li2016gan}, which degrades crystalline quality of the material and consequently the device performance. To overcome this problem, different growth techniques have been employed in the past, among which use of nanostructures is efficient and cost effective\cite{feng2017iii,himwas2015iii}. 
\par
Various morphological  nanostructures of III-nitride (especially GaN) have been reported \cite{kuykendall2003metalorganic,zhou2003physical,colby2010dislocation}, out of which the porous structure of GaN has shown  great potential for various applications\cite{soh2013nanopore,yam2007schottky,foll2006porous}. Most of the  growth of porous GaN is achieved by top down approaches such as chemical etching\cite{wang2004fabrication} or ion bombardment \cite{lian2006simultaneous}, which limit the device performance due to contamination, undesired interface and defect states, and degradation in crystallinity and composition. Previously, we have shown that by controlling V/III ratio, spontaneous formation of porous nanostructures can be achieved by using PA-MBE system,\cite{thakur2015surface,nayak2017nanostructuring,kesaria2011nitrogen,thakur2015electronic,kesaria2011nitrogen} which was followed by a few groups \cite{zhong2012growth,zhong2013characterization,poppitz2014microstructure,kushvaha2017quantum}. Investigations of the growth process of NwN  are highly important to control shape and size of the pores, to enhance light extraction efficiency\cite{nayak2017nanostructuring}. Poppitz \textit{et al.} \cite{poppitz2014microstructure} carried out a thickness dependent study of porous GaN  on 6H-SiC, where they found that the islands laterally elongate to form a network like structure. Further, Wu \textit{et al.}\cite{wu2009effects} demonstrated a similar pathway for the growth of ZnO nanowall grown on Sapphire substrate by Metal-Organic Chemical Vapour Deposition (MOCVD). Both these reports show step-wise evolution of the morphology, but no clear mechanism was proposed in either case. We intend to monitor the shape transition of the NwN that occur during the initial stage of the growth as it plays a crucial role in determining the final morphology of thin films and nanostructures\cite{rastelli2001reversible,barth2005engineering,tersoff1993shape,brune1998self,consonni2010nucleation}. In this report, we elucidate on growth mechanism of GaN NwN on Sapphire substrate by experimentally monitoring the evolution of surface morphology at intermediate stages of growth, complementing it by using \textit{first-principles} Density Functional Theory (DFT) simulations. We find that the shape transition for this unique nanostructure of GaN from intial 3D island is driven by Structure Dependent Attachment Kinetics (SDAK) induced instability a.k.a edge sharpening instability (ESI), which was previously used to understand the growth of snow crystal by Libbrecht \textit{et. al}\cite{libbrecht2011observations,libbrecht2012edge,libbrecht2017physical}. We have undertaken this complementary study of experiment and calculations to correlate such instability induced growth mechanism of thin films and/or nanostructures.

\section{Methods}
\subsection{Experimental Details}
The GaN films were grown under nitrogen rich conditions by using radio frequency Plasma Assisted Molecular Beam Epitaxy system (RF-PAMBE, SVTA-USA)  over bare c-plane of Sapphire with base pressure of $\approx$3$\times$10$^{-11}$ Torr. The detailed procedure of substrate preparation can be found elsewhere\cite{kesaria2011nitrogen}. Substrate temperature of 630 $^o$C and plasma forward power of 375W were maintained for all samples grown for this work. The other growth parameters are listed in Table \ref{tab}. 
The film structure was monitored \textit{in-situ} by reflection high energy electron diffraction (RHEED) with an acceleration voltage of 7 kV, the morphology was determined \textit{ex-situ} by field emission scanning electron microscope (FESEM) with an acceleration  voltage of 20 kV and atomic force microscopy (AFM) in contact mode.

\begin{table}[t]
 \caption{\label{tab}  Growth parameters} 
\begin{ruledtabular}
   \centering
 \begin{tabular}{ ccccccccc rrrrrrrrrrr }
 {Sample}  & {Ga-K cell}   & {N$_2$ flow}   & {Duration}   \\
 {Name}  & {temp ($^o$C)}  & {rate (sccm)}   & {(minute)}  \\
  \hline 
{A} & {1030} &{4.5} & {10}   \\
{B} & {1030} &{4.5} & {20}    \\
{C} & {1030} &{4.5} & {40}     \\
{D} & {1030} &{4.5} & {60}      \\
{E} & {1030} &{4.5} & {80}\\
 
\end{tabular}
\end{ruledtabular}
\end{table}
\subsection{Simulation Details}
Adsorption energy of Ga ad-atom on (20$\overline2$1) surface is estimated by total energy calculation using \textit{first-principles}  Density Functional Theory (DFT), as implemented in the SIESTA code\cite{soler2002siesta}. Generalized Gradient Approximation (GGA) proposed by Perdew \textit{et al.}\cite{perdew1996generalized} is used for the exchange and correlation function. We use the norm- conserving pseudopotentials of Troullier and Martins, with  valence electron configurations of Gallium, Nitrogen, and Hydrogen as 3d$^{10}$ 4s$^2$ 4p$^1$, 2s$^2$ 2p$^3$ and 1s$^1$, respectively. A double zeta  basis set with polarization functions is used for all atoms. Hartree and exchange correlation energies are evaluated on a uniform real-space grid of points with a defined maximum kinetic energy of 200 Ry. Brillouin Zone of w-GaN is sampled on a $\Gamma$- centered 5$\times$5$\times$3 mesh of k-points in the unit cell of reciprocal space\cite{pack1977special}. Positions of all the atoms are allowed to relax by the conjugate gradient technique to optimize energy until forces on each atom is less than 0.04 eV/\AA. The optimized lattice parameters of the unit cell are ‘a’ = 3.25 \AA \space and ‘c’ = 5.23 \AA. We construct a replica of the edge of the pyramid embedded with (20$\overline{2}$1) surface, which makes an angle 75$^o$ with c-plane. A vacuum of 15 \AA \space is added along all three axes to minimize the interaction of image configuration. The bottom layer and surface atoms, except the ones at (20$\overline{2}$1) facets, were passivated with Hydrogen (see Fig. \ref{model}(c)). The adsorption energy of Ga on different sites of (20$\overline{2}$1) surface is estimated by the equation
\begin{equation*}
\mathrm{E_{ads}(Ga) = E_{tot}(adatom+wedge)-E_{tot}(wedge)-\mu_{Ga}}
\end{equation*}
where $\mathrm{E_{tot} (adatom+wedge)}$ is the total energy of the combined system with Ga ad-atoms and
the wedge shape, $\mathrm{E_{tot} (wedge)}$, is the total energy of the pristine wedge and $\mu_\mathrm{Ga}$ is $\dfrac{1}{8} ^{th}$ the total energy of $\alpha$-Ga. The potential energy surface (PES) is constructed by placing Ga ad-atoms at various sites and allowing them to relax along the perpendicular to the (20$\overline{2}$1) surface.
\section{Results}
\begin{figure}[!htb]
    \centering
    \includegraphics[width=16cm]{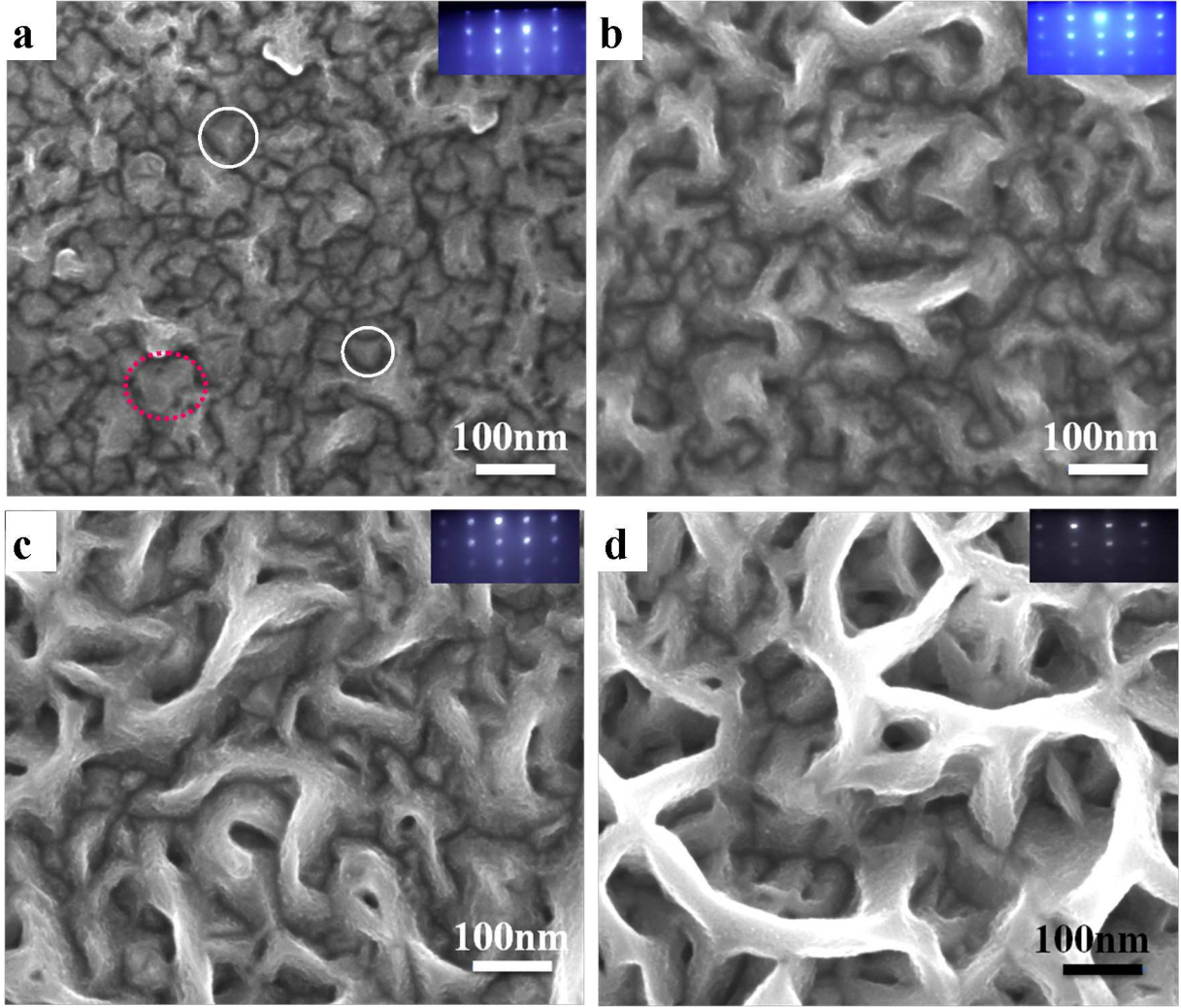}
    \caption{(a), (b), (c), (d) show FESEM images of sample B, C, D and E, respectively. Inset shows RHEED pattern of respective samples.}
    \label{sem}
\end{figure}
\begin{figure}[!htb]
    \centering
    \includegraphics[width=16cm]{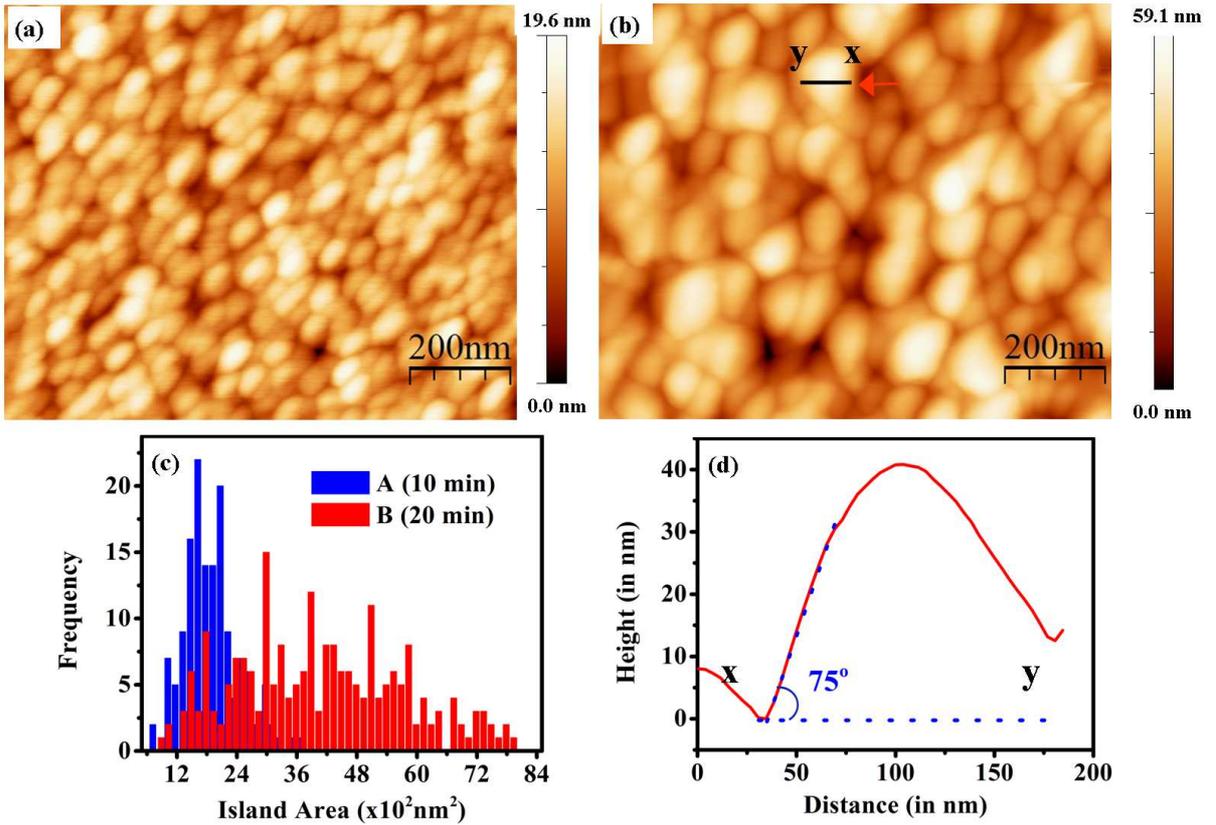}
    \caption{(a) and (b) show AFM image recorded in contact mode of samples A and B, respectively.
Fig. (c) shows size distribution (area) of islands of sample A (blue bars) and B (red bars). Fig.(d) shows line scan pattern of the island (x,y) shown by a red arrow.}
    \label{afm}
\end{figure}

To look at the initial nucleation  and evolution of surface morphology, we carry out growth of GaN NwN for durations of 10, 20, 40, 60 and 80 minutes. The corresponding FESEM images of the resulting morphology are shown in Fig.\ref{sem}. A careful visualization of the FESEM image of sample B shows formation of tetrahedron shaped faceted island structures (3 faced pyramids) (see inside the white circle of Fig.\ref{sem} (a)). Along with these individual islands, we also observe a few Y-shaped structures (see inside red circle of Fig.\ref{sem} (a)).  Sample C, grown for higher duration (40 min) shows mostly Y-shaped structures with a few tetrahedron islands. Further, the length of tails of Y-shaped structure for sample B are $\approx$ 45-55 nm while for sample C they are longer ($\approx$ 55-75 nm). Such increase in length of the tail of the Y-shaped structure with increased growth durations suggests that enhanced growth of the edges is responsible for the Y-shaped morphology. However, not all  Y-shaped islands are connected to each other in both samples B and C. For samples grown for higher duration (samples D and E), we observe the Y- shaped structures connect with each other to  form wall like features. To further look at earlier stage of island formation, we grow a film for 10 minutes under identical growth conditions. Fig.\ref{afm} (a) and (b) shows the AFM image of  samples A and B, respectively. At this early stage of growth (10 mins), we observe the formation of mostly oval shaped 3D islands with a large density ($\approx$ 3.24 $\times 10^{10}$ cm$^{-2}$, see Fig.\ref{afm} (a)). The island size distribution is plotted in Fig.\ref{afm}(c) where we find that the island sizes have a relatively uniform distribution with mean area $\approx$ 1900 nm$^2$  with a standard deviation of 550 nm$^2$. However, with increase in the deposition time to 20 minutes, size of most islands increases while that of a few ($\approx$ 1.05$\times 10^{10}$ cm$^{-2}$) remains unchanged (see Fig.\ref{afm}(c)), which results in a broad distribution of island sizes ($\approx$ 1200-8000 nm$^2$).  It is very clear from SEM and AFM images, that the in-plane growth along the edges of the tetrahedron is dominant, resulting in the formation of  the Y-shaped structure. To understand such edge enhanced growth mechanism, it is necessary to understand the dynamics of adatoms on the side facets of the tetrahedron. From line-scan analysis of  selected islands (see Fig.\ref{afm} (d)) we find that the side walls of the tetrahedron, observed in sample B, make an angle of  75 $\pm$ 1$^o$ with c-plane, which is identified as (20$\overline{2}$1) surface of GaN.
 \par
\begin{figure}[!htb]
    \centering
    \includegraphics[width=16cm]{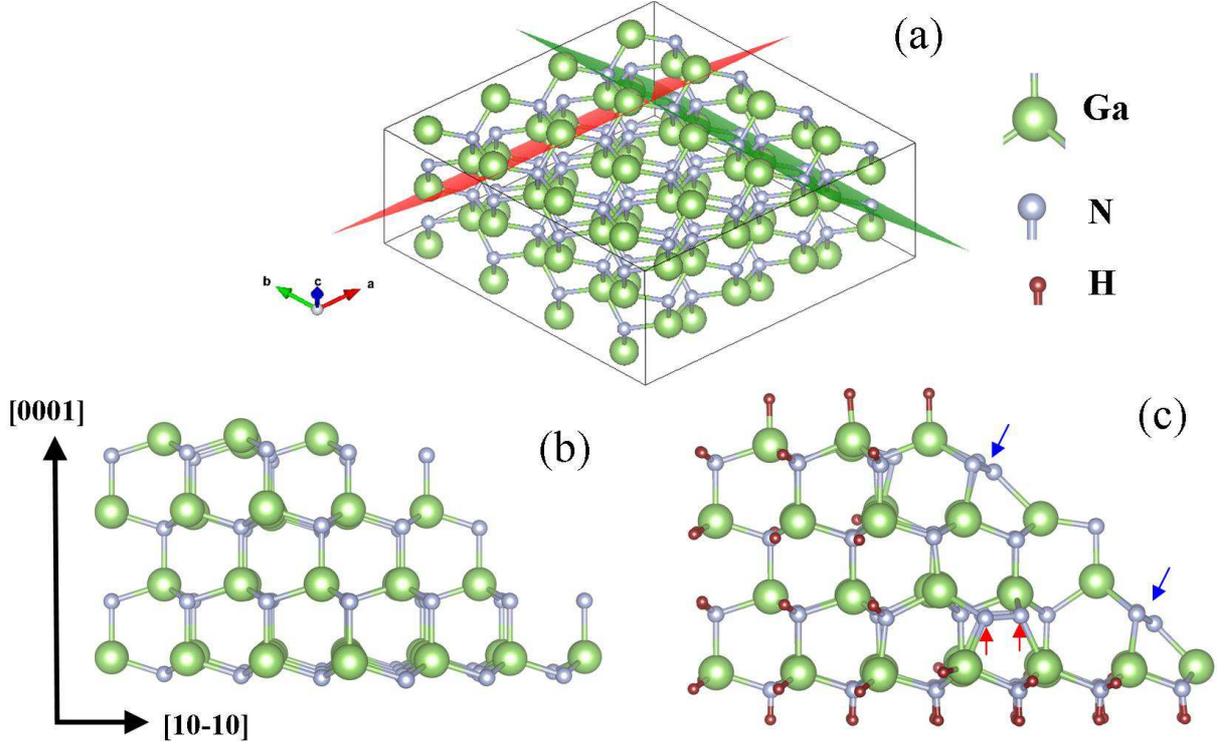}
    \caption{(a) shows the construction of the wedge shaped structure. Both red and green surfaces
represents \{20$\overline{2}$1\} planes. Fig.(b) and (c) shows unrelaxed and relaxed structures of the wedge
shape.}
    \label{model}
\end{figure}
\begin{figure}[!htb]
    \centering
    \includegraphics[width=16cm]{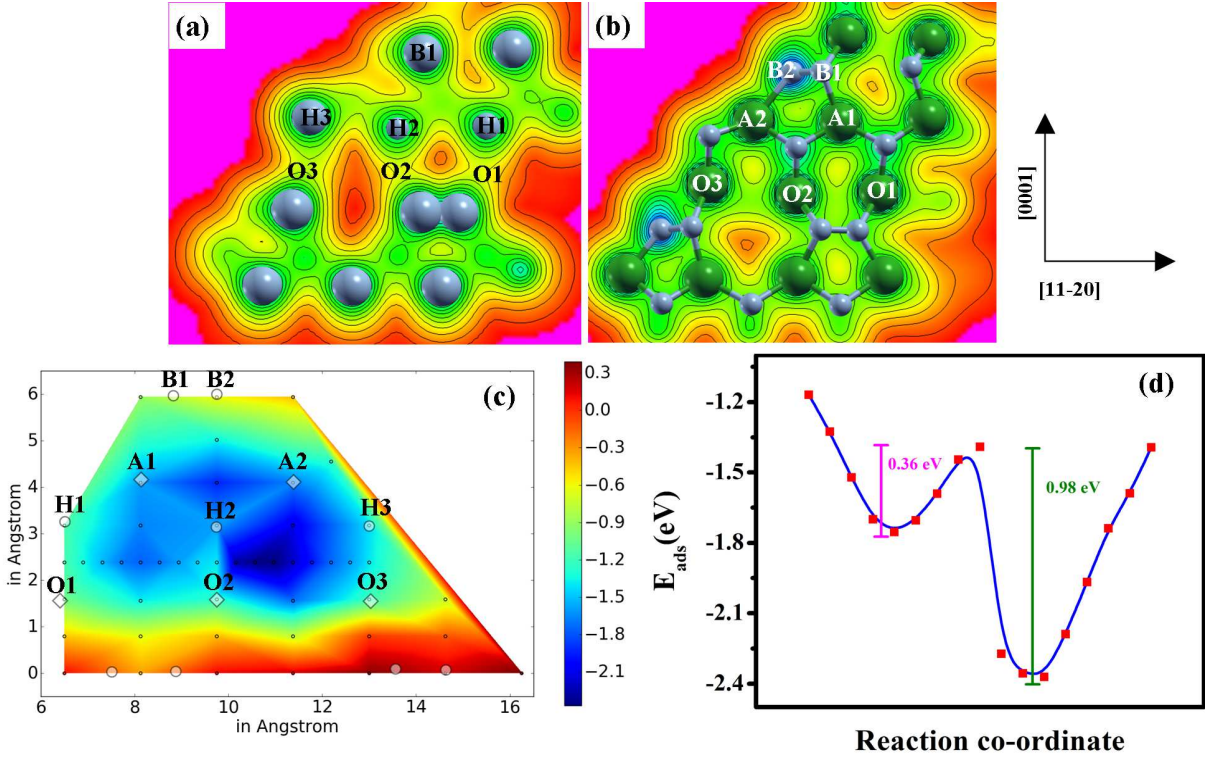}
    \caption{(a) and (b) show surface electron charge density plots of 1$^{st}$ and 2$^{nd}$ atomic layers of the (20$\overline{2}$1)surface which constitute the side surfaces of the tetrahedron shaped island. Fig.(c) shows contour plot of the estimated adsorption energy of Ga ad-atom at various sites on the surface. Fig.(d) depicts the
adsorption energy as a function of spatial co-ordinates along [11$\overline{2}$0].}
    \label{ads}
\end{figure}

 Thus, to study the surface diffusion of Ga adatoms on the side surface of the tetrahedron, we construct a wedge shaped structure, where the edge is the intersection  of \{20$\overline{2}$1\} planes and is shown in Fig.\ref{model}. 
In the relaxed structure of the wedge, we find the Nitrogen (N) having two dangling bonds at surface forms  dimers (see Fig.\ref{model} (b) and (c)), whereas at the edge a trimer is formed (see blue arrows in Fig.\ref{model}(c)) due to Stirling's instability\cite{lee2011dimer}. We estimate the adsorption energy of Ga ad-atoms at various positions of the  relaxed structure; on
 the surface and at the edge (see Fig.\ref{ads}(c)). We find that the most favorable sites for Ga adatom are the
 hollow sites (see Fig.\ref{ads}(c)).  The adsorption energy of Ga adatom at the hollow site
 of O1-O2-A1 triangle is -1.75 eV where the width of the wedge is 6.35 \AA \space whereas, for the
 hollow site of O2-O3-A2 triangle the adsorption energy is -2.37 eV, while the width is 3.14 \AA \space. We estimate the adsorption energy at various equivalent points on the surface such as at O1, O2 and O3, H1, H2
 and H3, as well as at A1 and A2 (see Fig.\ref{ads} (b) and (c)) and find  that the  adsorption
 energy reduces with reduction in the dimension of the wedge. However, interestingly, nearby the N-N dimers and N-N-N trimers, the adsorption energy is substantially high which may be due to the strong bonding character of N atoms. Further, to look at the diffusion barrier of the Ga
 adatom on the (20$\overline{2}$1) facets, we estimate the adsorption energy at very close distances (0.4 \AA) along the [11$\overline{2}$0] direction (see Fig. \ref{ads} (d)). It is very clear that, adatoms have to overcome a barrier potential of nearly 0.36 eV to cross from hollow site of O1-O2-A1 triangle to hollow site of O2-O3-A2 triangle, whereas for reverse diffusion the barrier potential is  0.98 eV.

\section{Discussion}
The nucleation mechanism of GaN on Sapphire has been studied widely in the literature\cite{lukin2012nucleation,degave2002analysis,richards1995atomic}. Degave \textit{et al.}\cite{degave2002analysis} show that growth of GaN on bare surface of Sapphire results in the formation of  3D islands from the early stages of the growth, which is consistent with our observation of the morphology of sample A. It is well known that, in highly lattice mismatched systems (e.g. GaN on c-Sapphire (16\%)) the growth mode is governed by the interface and surface energies only\cite{shchukin1999spontaneous}. In such systems  the strain relaxation of grown island occurs in two different relaxation regimes. At the early stage of growth the elastic relaxation is preferable where the nuclei evolves by aiming to minimize total free energy per unit volume by relieving the lattice mismatch induced strain ($\Delta E_{elastic} < 0$) and accordingly by reducing their stored elastic strain energy and/or the nuclei attain the lower energy state by changing its shape (such as formation of crystal facets), surface area ($\Delta A > 0$)\cite{vanderbilt1990elastic}, while with increase in film thickness the plastic relaxation occurs  by formation of misfit dislocations\cite{herman2013epitaxy}.
\par
\begin{figure}
    \centering
    \includegraphics[width=10cm]{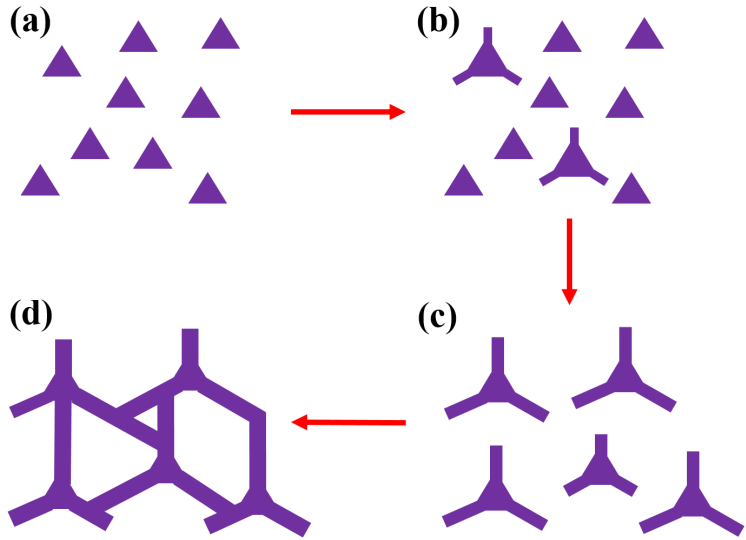}
    \caption{show schematic of the shape transition and evolution of surface morphology of GaN NwN. }
    \label{evolution}
\end{figure}

 Further, as observed experimentally the wider distribution of island sizes in sample B may be due to the fact that
the growth of the dislocated islands is energetically preferable than coherent islands, since the  strain energy per unit volume is minimum in the former while later leads to increase in strain energy \cite{krishnamurthy1991microstructural,drucker1993coherent}, which suggest that the plastic relaxation mechanism took place before 20 minutes growth of the samples . The appearance of pyramidal shaped islands is possible due to the difference in adatoms attachment and/or site exchange rates between atomic steps induced by  Ga adlayers on GaN (0001)\cite{fang2008silicon}. The shape transition of such pyramidal islands is mainly governed by the interplay between stored elastic strain energy, total interface and surface energies of all facets composing of all the surfaces of the pyramid. Shchukin \textit{et al.}\cite{shchukin1999spontaneous} pointed out that, edges of the two surfaces could also play a crucial role in the final shape transition, as at this point the discontinuity of the surface stress occurs. Further, Libbrecht \textit{et al.}\cite{libbrecht2012edge} attributed  the edge enhanced growth, as seen here, is due to the structure dependent attachment kinetics (SDAK) instability that becomes dominant when diffusion related growth is coupled with structure dependent attachment kinetics of adatoms.  Typically, SDAK instability is the consequence of decrease in the nucleation barrier on a faceted surface when the width of the facets decreases, which is consistent from our \textit{first-principles} calculations.  It is relatively easy for Ga ad-atoms to diffuse from  thicker part of the wedge to relatively thinner part (towards edge) but the diffusion of  ad-atoms from edges of the wedge to the thicker part is energetically not preferable as it has to overcome a large potential barrier of 0.98 eV. Thus, it is evident that a large amount of flow of ad-atoms towards the edges of the pyramid is responsible for edge enhanced growth. This anisotropic growth further sharpens the edge, which enhances the edge growth rate. This positive feedback results in a growth instability which promotes formation of sharp edges. This process continues  to lengthen the edges of the tetrahedron until it meets with other similar growing edge of another proximal island to form a junction. Once such junctions are formed, the in-plane growth of the edge stops due to unavailability of thin edges for attachment of the adatoms and the dominant growth will now be along perpendicular direction to the surface of the substrate. Based on the FESEM  images and results from \textit{first-principles} calculations we present a 2D schematics on the morphological evolution and the shape transition of NwN from 3D islands in Fig.\ref{evolution}. Further, with increase in the growth time, the side facets of the tetrahedron change from (20$\overline {2}$1)  to (10$\overline {1}$0) surface \cite{nayak2017nanostructuring} due to low surface free energy of the later\cite{dreyer2014absolute}.  The final surface morphology observed with higher growth time\cite{thakur2015surface,nayak2017nanostructuring,kesaria2011nitrogen,thakur2015electronic,kesaria2011nitrogen} is  consistent with the morphology (hollow columnar) proposed by Libbrecht \textit{et al.} \cite{libbrecht2012edge}.  
As  discussed earlier, the adsorption energy of Ga adatoms  nearby the N-N dimers and N-N-N trimers (including at edge) is substantially high, due to the strong bonding character of N atoms. Despite adding an adatom nearby to the dimers they maintain their stability. Such surface and/or edge reconstruction of N atoms are commonly observed in different semipolar surfaces of III-V semiconductors, but their desorption as N$_2$ molecule is  energetically favorable\cite{lee2011dimer,nishinaga2014handbook}. Thus, the possibility of the presence of dimers and trimers in the grown crystals is negligible.

\section{Summary}
In summary, we have carried out growth of GaN NwN for various durations to monitor the evolution of the surface morphology. At the initial stages we observe the formation of tetrahedron shaped 3D island. With increase in growth time, we observe a wider distribution of islands sizes  due to co-existence of coherent and dislocation mediated islands.  From \textit{first-principles} simulations we find that the thinner edges of such tetrahedron are more favorable for attachment of adatoms than their thicker counterparts, resulting in edge driven growth. We find an anisotropy in the diffusion barrier for adatoms on (20$\overline{2}$1) surface depending upon width of the facets, which leads to the formation of the nanowall network. We infer that  the evolution from 3D island to NwN morphology is a clear example of SDAK instabilities induced shape transition in thin film growth.  

\section*{acknowledgement}
The authors thank Professor C. N. R. Rao for his support and guidance.  SKN acknowledges DST, Govt. of India, for  SRF and RK acknowledges UGC, Govt. of India for  JRF. All authors acknowledge JNCASR for facilities.

\bibliographystyle{apsrev4-1}
\bibliography{model} 
\end{document}